\documentclass{ws-rv961x669}
\usepackage{subfigure}   
\usepackage{ws-rv-thm}   
\usepackage{ws-rv-van}   
\usepackage{ws-index}   
\makeindex

\usepackage{graphicx}
\usepackage{subfigure}
\usepackage{textcomp}
\usepackage{slashed}
\usepackage{wrapfig}

\def\beq{\begin{equation}}
\def\eeq{\end{equation}}
\def\theta{\vartheta}

\newcommand{\ba}{\begin{eqnarray}}
\newcommand{\ea}{\end{eqnarray}}
\newcommand{\lsim}   {\mathrel{\mathop{\kern 0pt \rlap
  {\raise.2ex\hbox{$<$}}}
  \lower.9ex\hbox{\kern-.190em $\sim$}}}
\newcommand{\gsim}   {\mathrel{\mathop{\kern 0pt \rlap
  {\raise.2ex\hbox{$>$}}}
  \lower.9ex\hbox{\kern-.190em $\sim$}}}

\begin{document}



\title{Neutrino Physics and Astrophysics}

\chapter[Cosmological Neutrinos]{Cosmological Neutrinos*}

\author[Floyd W. Stecker]{Floyd W. Stecker\footnote{Floyd.W.Stecker@nasa.gov}}
\address{NASA Goddard Space Flight Center \\ Greenbelt, MD 20771 USA \\
and University of California Los Angeles \\ Los Angeles, CA 90095} 

\vspace{1.0cm}
 
\noindent     {\it In the beginning...there was light.}
                   \\\ \\ -- Genesis, 1:1-1:3

\vspace{1cm}

\begin{abstract}

\noindent Within the context of hot big-bang cosmology, a cosmic background of  
presently low energy neutrinos is predicted to exist in concert with the photons of
the cosmic background radiation. The number density of the cosmological neutrinos
is of the same order as that of the photons of the cosmic background radiation.
That makes neutrinos the second most abundant particle species in the universe.
In the early universe, when these neutrinos were highly relativistic, their effects in
determining the ultimate structure and evolution of the universe were significant.

 
\end{abstract}

\vspace{1.5in}

*To be published in "Neutrino Physics and Astrophysics", edited by F. W. Stecker, in Encyclopedia of Cosmology II, edited by G. G. Fazio, World Scientific Publishing Company, Singapore, 2022.

\newpage

\tableofcontents  

\newpage
   
\vspace{0.2cm}

~
\section{Introduction}
~   
In Chapter 1 we outlined the history of neutrino physics and astrophysics and discussed the
present state of neutrino mass measurements. This chapter will present a primer on the
{\index~cosmological neutrino background} that is a relic of the {\index~hot big bang}. We
briefly summarize the implications of the cosmological neutrino background on neutrino physics
and {\it vice versa}. The predicted {\index~light element production} in the early big-bang, assuming standard {\index~weak interaction} neutrino physics, is in excellent agreement observations of the {\index~cosmic abundances} of helium, deuterium and lithium in the early universe. Therefore, we will not consider the effects of more exotic physics in this chapter.

\begin{figure}
    \centering
    \includegraphics[width=0.7\textwidth]{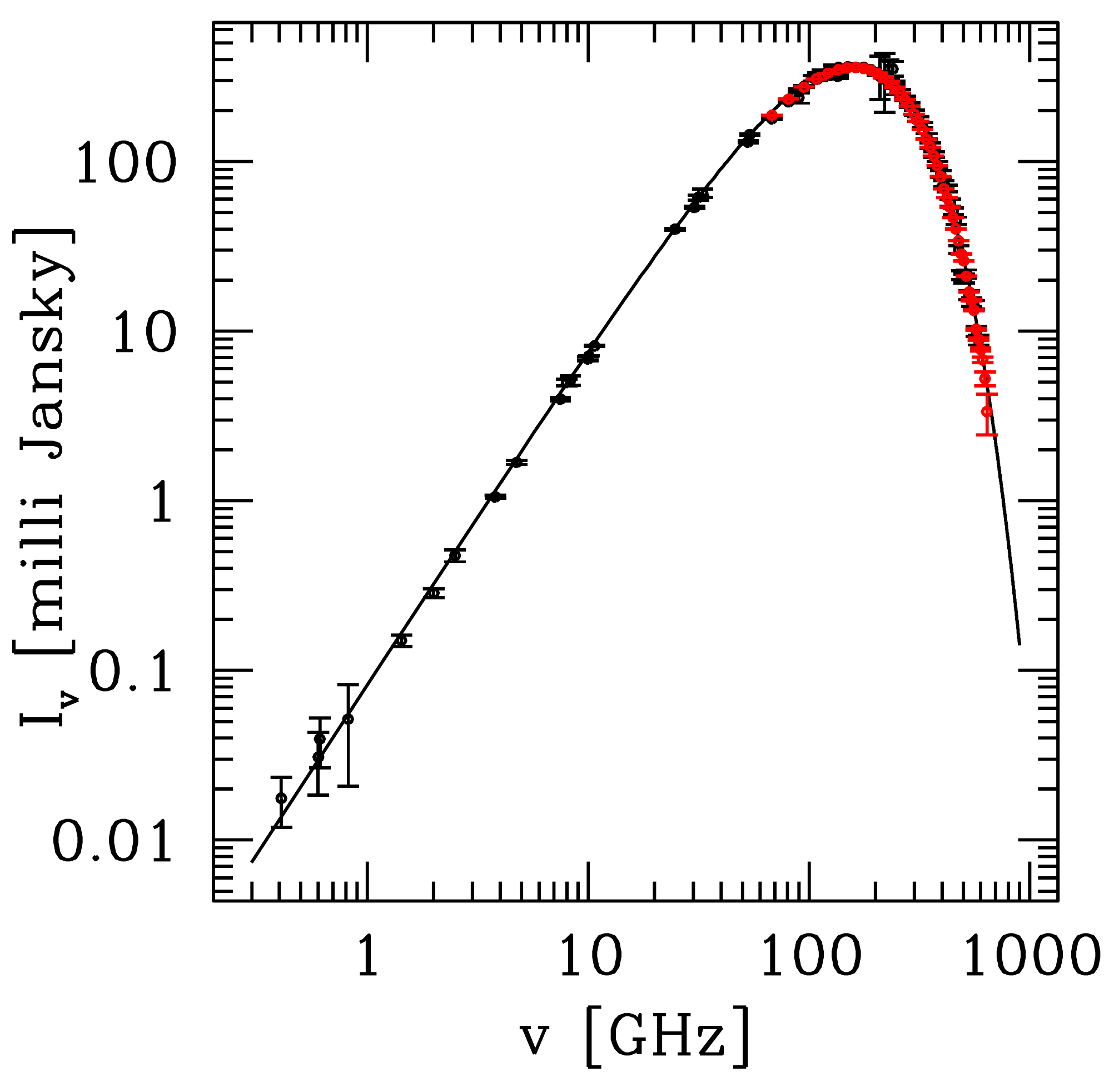}
    \caption{The blackbody spectrum of the {\index~cosmic background radiation} as first measured by the {\index~Cosmic Background Explorer (COBE)}~\cite{Mather:1990tfx}. The red points around the peak
    were measured by the diffuse infrared background explorer (DIRBE) on the COBE satellite~\cite{Fixsen:1996nj
    }. Figure courtesy of Ruth Durrer.~\cite{Durrer:2015lza}}
    \label{cmb}
\end{figure}

The cosmic microwave background radiation was first detected in 1966 by Penzias and Wilson~\cite{Penzias:1965wn}
and elucidated by R.H. Dicke and P.J.E. Peebles and P.G. Roll, and D.T. Wilkinson~\cite{Dicke:1965zz}.
At a meeting of the American Physical Society in 1990, to a spontaneous standing ovation,
{\index~John Mather}, the leader of the COBE group, announced their observational results showing a full-blown perfect
2.73 K black body spectrum of the cosmic background radiation (CBR) as shown in Figure \ref{cmb}. 
This was solid proof of the hot big-bang picture
of the evolution of the universe~\cite{Mather:1990tfx}. In this picture, the early universe began in a hot, dense state
that expanded and cooled adiabatically with time, explaining the cosmological origin and abundances of helium, deuterium,
and lithium. An implication of this scenario was the predicted existence of a presently ultracold 1.95 K cosmic neutrino background coexisting with the background of cosmic blackbody radiation.

\section{The cosmic background radiation (CBR) and the cosmic neutrino background (C$\nu$B)}


In the hot big-bang scenario, at a time when $T > T_{dec}$, where $T_{dec}$, is the {\index~decoupling temperature} (see below),   the "radiation" in the universe, consisted of photons, electrons, positrons, and neutrinos, that were all in thermal equilibrium. 
At that time, the ratio of neutrino density to photon density in  thermal equilibrium was given by
\begin{equation}
\frac{n_{\nu}}{n_\gamma} = \frac{\omega_\nu}{\omega_\gamma}\frac{{\int_0^\infty} {F(\epsilon_\nu) d\epsilon_\nu }} {{\int_0^\infty} {B(\epsilon_\gamma) d\epsilon_\gamma}} = {\frac{3}{4}} 
\label{densfrac} 
\end{equation}

\noindent where $ \omega_\gamma = 2$ is the number of photon degrees of freedom, $\omega_\nu$ is the number of neutrino
degrees of freedom, $f$, taken to be 2 per flavor that were in thermal equilibrium when $T \sim 1$ MeV (only $\nu_L$
and $\bar\nu_R$ meet this criterion). Thus, the relation (\ref{densfrac}) is independent of the {\index~Dirac} or {\index~Majorana} nature of the neutrino. Here, $F(\epsilon_\nu)$ and $B(\epsilon_\gamma)$ designate the {\index~Fermi-Dirac distribution} and {\index~Planck blackbody distribution} respectively. The factor of 3/4 comes from the ratio of the integrals of the Fermi-Dirac distribution and the Planck distribution.\footnote{In this discussion, we assume that the neutrino chemical potential, $\mu_\nu$, that could generally appear in $F(\epsilon_\nu)$, is equal to zero. This is consistent with the very strong limits on $\mu_{\nu_e}$. We note that the strong bounds on $\mu_{\nu_e}$ apply to all flavors, since neutrino oscillations lead to approximate flavor equilibrium before big-bang nucleosynthesis (BBN) occurs. Thus, $\mu_{\nu_f} = 0$ is consistent with the agreement between the BBN predictions and observations. We also take the effective number of neutrinos to be three, also consistent with observations.} 

The decoupling temperature, $T_{dec}$, when the neutrinos fell out of thermal equilibrium, occurred
when the $\nu e$ weak interaction rate was equal to the expansion rate of the universe.\footnote{At $T \sim$ 3 MeV
neutrinos have already gone out of equilibrium with electrons because the cross section for $\nu_e n \rightarrow
e^- p$ is smaller than that for $e^+e^- \rightarrow \gamma\gamma$.}
  
Thus, an estimate of the decoupling temperature can be found by equating the thermally
averaged value of the $\nu_e e$  {\index~weak interaction rate}, $\Gamma_{\nu_e e}$, and the {\index~expansion rate,
$H$}.

It follows from the weak interaction theory of Fermi that for $T < \sim100$ GeV,
\vspace{10pt}
\begin{equation}
\Gamma_{\nu e} \propto  n_\nu G_F^2 T^5\,\, ,
\label{Gamma}
\vspace{10pt}
\end{equation}
%
where $n_\nu$ is the {\index~neutrino number density} and $G_F^2 = g^4/(32M_{W}^4)$ is the square of the {\index~Fermi constant}, with $g$ being the {\index~weak coupling constant} and $M_{W} = 80 \ {\rm GeV}$ being the {\index~$W$-boson mass} in {\index~electroweak theory}.
 
The expansion rate ({\index~Hubble parameter}), $H$,
is given by
\vspace{10pt}
\begin{equation}
H = \sqrt{\frac{8\pi \rho}{3M_P^2}} \,\, ,
\label{H}
\vspace{10pt}
\end{equation}
%
where $\rho$ is the total energy density and $M_P = \sqrt{\hbar c/G_N}$ is the {\index~Planck mass}
with $G_N$ being Newton's constant. 

In the {\index~radiation dominated era} $\rho$ is given by

\begin{equation}
\rho \ \simeq \ \rho_{\rm rad} \ = \ {\pi^2 \over 30} \left( 2 + {7 \over 2} + {7 \over 4}f \right) T^4 ,
\label{rho}
\vspace{10pt}
\end{equation}
given in terms of the {\index~number of neutrino flavors}, $f$\footnote{We assume here the non-existence of sterile neutrinos}. Equations (\ref{H}) and (\ref{rho}) show that {\it the 
expansion rate of the early universe is partially determined by the number of neutrino flavors, $f$.}

In natural units ($\hbar = c = 1$), 
$M_P = G^{-1/2}$ This gives $H \simeq T^2/M_P$.
By roughly equating equations (\ref{Gamma}) and (\ref{H})
we find that the temperature of the universe at decoupling was, $T_{\rm dec} \simeq 1$ MeV. 

Shortly after the neutrinos decoupled from the radiation field, when the photon temperature dropped below
the electron mass, i.e., $T \simeq m_e \simeq 1/2$ MeV
this enabled irreversible $e^+e^-$ annihilation via $e^+ e^- \rightarrow 2\gamma$ to occur, 
with the energy release going into the CBR, raising its temperature. One can assume 
that this entropy transfer did not affect the temperature of the
neutrinos because they were already completely decoupled.
Therefore, when $T < m_e$, additional photons were created by $e^+e^-$ annihilation. Thus, a new factor multiplying the
photon number density is determined by the additional entropy per unit volume added to the photon component, viz.,
11/3. Thus, for $T < m_e$, 
\vspace{10pt}
\begin{equation}
{{n_\nu}\over{n_\gamma}} = {{3}\over{11}}f.
\label{nun}
\vspace{10pt}
\end{equation}

From that point on, the ratio between the temperatures of relic photons
and neutrinos became $T_\gamma/T_\nu=(11/4)^{1/3}\simeq 1.4$, as follows from equations (\ref{densfrac}) and (\ref{nun}).
Taking the present value of the CBR $T_{0} = 2.73$ K, this relation gives $T_{\nu,0} = 1.95$ K.  

For almost all of its history the cosmic photon background radiation field expanded isotropically with the mean photon
with redshift, $z$. This leads to the redshift-dependent relations:
\ba
T_{\gamma} = T_{0}(1 + z), \\
<\epsilon> = 2.7kT = 2.7kT_{0}(1 + z), \\
n(\gamma) = n_0(1 + z)^3, \\
\rho(\gamma) = <\epsilon>n(\gamma) = \rho_{0}(1 + z)^4.
\ea

It follows from equation (\ref{nun}) that the neutrino number density per flavor 
is determined by the temperature, $T_{\nu}$
\begin{equation}
n_{\nu} = \frac{3}{11}\;n_\gamma =
\frac{6\zeta(3)}{11\pi^2}\;T_\gamma^3~,
\label{nunumber}
\end{equation}

This further leads to the relation
\begin{equation}
\rho_\nu (m_\nu \ll T_\nu) 
\frac{7\pi^2}{120}
\left(\frac{4}{11}\right)^{4/3}\;T_\gamma^4~.
\end{equation}

In the non-relativistic limit, $\rho_\nu (m_\nu \gg T_\nu)  =  m_\nu n_\nu$,
so that the contribution of massive neutrinos to the energy
density in the non-relativistic limit is a function of the mass (or
the sum of masses of all neutrino states, given $\Sigma m_i \gg T_\nu$).

At the present time the {\index~CMB temperature} is measured to be $T_0 = $2.73 K corresponding to a photon density $n_\gamma \simeq 4 \times 10^8$ m$^{-3}$. Thus, from equation (\ref{nun}), $n_\nu \simeq 3.4 \times 10^8$ m$^{-3}$, i.e., more than one neutrino for each man, woman and child in the United States in a single cubic meter!

\section{Neutrino mass and cosmological mass density}

Let us begin by giving the usual definitions relating
the Hubble parameter to the overall mass density in
the universe. The {\index~Hubble parameter} relates the average
{\index~expansion velocity} of a point in space such as a
galaxy to its distance. Locally, it is designated by 
the constant, $H_0$. While there is some controversy about its exact
value, depending on the measurement technique used~\cite{Verde:2019ivm}, we take it here to
to be $H_0 \sim$ 70 km/s/Mpc. We also define $h = H_0$ /(100 km/s/Mpc)
and take $h \simeq$ 0.7.

The {\index~critical mass density} needed to gravitationally close the universe with zero
curvature is given by
\vspace{10pt}
\begin{equation}
\rho_c = {{3H_0^2}\over{8\pi G_N}}\,
\vspace{10pt}
\end{equation}
$H_0$ being the Hubble parameter and $G_N$ being the gravitational constant. 

Because neutrinos have mass, cosmic background neutrinos are a form of dark matter.
Thus, they contribute to the overall mass density of the
universe. Defining the ratios $\Omega_m \equiv \rho_m/\rho_c$, where $\rho_m$
is the total mass density of the universe, we can also define the contribution
of cosmological neutrinos to the mass density of the universe.
If we define the parameters $\Omega_\nu = \rho_\nu/\rho_c$ and $h = H_0$ in units
of 100 km/s/Mpc,
the present contribution to the matter density of $f_\nu$ neutrino
species with standard weak interactions is given by
\vspace{10pt}
\begin{equation}
\Omega_\nu h^2 = f_\nu \frac{<m_\nu>}{93.8 \, {\rm eV}}
\label{omega}
\vspace{10pt}
\end{equation}
where $<m_\nu>$ is the average mass over the number of 
neutrino flavors, $f_\nu$.

Neutrino oscillation observations shed some light on neutrino masses. They indeed prove that neutrinos
have masses. However, the {\index~neutrino oscillation} periods are determined by parameters involving the differences between the squares of the neutrino masses, e.g., ($m_{2}^{2}$ - $m_{1}^{2}$), so that the individual masses themselves are not determined by the oscillations (see Chapter 2). The oscillation phenomenon proves that neutrinos have mass eigenstates that are combinations of their flavor eigenstates and that neutrinos have non-zero masses. 

Although we do not know the masses of the neutrino mass eigenstates, we know that they are all much lighter than those of the other known particles. As of this writing, the {\index~Karlsruhe Tritium Neutrino experiment (KATRIN)}, by studying the endpoint energy of the electron energy spectrum from tritium $\beta$-decay, has placed an upper limit on the neutrino-mass scale of 1.1 eV at a 90\% confidence level~\cite{KATRIN:2019yun}. On the other hand, neutrino oscillation experiments have placed a {\it lower limit} of 0.06 eV on the sum of the neutrino masses~\cite{deSalas:2020pgw,RoyChoudhury:2019hls}. This mass range shows that $\Omega_\nu h^2 \ll 1$, as follows from equation (\ref{omega}). If we take $ <m_\nu> \sim 1$ eV and $f_\nu = 3$ in equation (\ref{omega}), we obtain a value for $\Omega_{\nu}$ of $\sim$ 0.03. However, although neutrinos cannot account for a large portion of the dark matter in the universe, {\it they are presently the only identified component of the {\index~dark matter}}.

\section{Cosmological Effects of Neutrinos}

\subsection{Big Bang Nucleosynthesis}

As we have seen, at present neutrinos have little effect on the mass density of the universe. However, in the early
radiation dominated universe, when they were relativistic they played a significant role in the dynamics of the 
universe. Their effects in determining the ultimate structure and evolution of the universe were significant.

It follows from equations (\ref{H}) and (\ref{rho}) that the expansion rate in the early universe is determined by
the number of neutrino flavors $f$. The abundances of primordial helium, deuterium, and lithium 
made in the first few minutes after the big bang are determined by this expansion rate~\cite{Bernstein:1988ad}. In particular, 
the abundance of helium is sensitive to the available abundance of neutrons which is in turn determined by the
$n/p$ ratio at the {\index~"freezeout" temperature} $T_{\rm f}$ when $\Gamma_{\nu e} > H$. In order to account for the observed abundances, one obtains a value of $f \simeq 3$, viz., the known number of neutrino flavors~\cite{Fields:2019pfx,Olive:2021noj}.
A more precise determination of the number of neutrino flavors comes from a determination of the mass width of the {\index~$Z$
boson} using the {\index~LEP accelerator} at CERN~\cite{Mele:2015etc}.

The following discussion therefore implicitly assumes $f = 3$ in equation (\ref{rho}):
\vspace{12pt}

Before $ T \sim 1$ MeV protons and neutrons are kept in thermal equilibrium by the weak interactions:
\vspace{10pt}
\begin{eqnarray}
\label{eq:weak}
{\rm n} + \nu_{\rm e} &\leftrightarrow& {\rm p} + {\rm e}^{-} \,, \\
{\rm n} + {\rm e}^{+} &\leftrightarrow& {\rm p} + \bar{\nu}_{\rm e} \,. 
\vspace{10pt}
\end{eqnarray}
The neutrons and protons are non-relativistic, i.e., $T \ll m_n ,m_p$) so they have a {\index~Maxwell-Boltzmann distribution} and their relative number densities are given by
\vspace{10pt}
\begin{equation}
\label{nnnp}
\frac{N_{\rm n}}{N_{\rm p}} 
\approx \exp{ \left[- \frac{ (m_{\rm n} - m_{\rm p}) }T \right]}  \,.
\vspace{10pt}
\end{equation}
As long as $T \gg  (m_{\rm n} - m_{\rm p}) = 1.3$ MeV, $N_{\rm n} \sim N_{\rm p}$. 
However, at a later time, when $T < (m_{\rm n} - m_{\rm p}$), then $N_{\rm n} < N_{\rm p}$.  A detailed calculation shows that once $T_{\rm fo} \sim 0.8 \, {\rm MeV}$, the abundance ratio $N_{\rm n}/N_{\rm p} \sim 0.2$.

The production of the nuclei of the light elements then occurs through the reaction sequence
chain of reactions:
\begin{eqnarray}
{\rm p}+ {\rm n} &\rightarrow& {\rm D} \,, \\
{\rm D} +{\rm p} &\rightarrow&  {}^3{\rm He} \,, \\
{\rm D} +{\rm D} &\rightarrow&  {}^4{\rm He} \,, ...
\vspace{10pt}
\end{eqnarray}
The destruction of nuclei by {\index~Wien tail} of the blackbody photon distribution effectively stops once the temperature is less than 
$\sim 0.1$ MeV. At this point the neutron to proton ratio drops to $N_{\rm n}/N_{\rm p} \simeq 0.18$. Then, most of the remaining neutrons form ${}^4 {\rm He}$. For a further discussion see Ref.~\refcite{Kolb:1990vq}.
 
\subsection{Blackbody Temperature Perturbations}

During the era of radiation dominance the gravitational effects induced by inhomogeneities in the photon and neutrino backgrounds were comparable. The faster {\index~cosmological expansion} owing to the neutrino background produces effects on the acoustic and damping angular scales of the cosmic microwave background. (See equation (\ref{rho})). Following decoupling ($T < T_{dec})$, the neutrinos free stream at a velocity $v_{\nu} \simeq c$. The gravitational effect of neutrino background perturbations suppresses the {\index~acoustic peaks} in the microwave background for the {\index~multipoles} with $l \gtrsim 200$ and enhances the amplitude of matter fluctuations on these scales.  

In addition, the perturbations of relativistic neutrinos generate a unique phase shift in the acoustic oscillations of the CBR that for adiabatic initial conditions cannot be caused by any other standard physics. The origin of the shift is an effect of the neutrino {\index~free-streaming} velocity exceeding the {\index~sound speed} of the photon-baryon plasma.~\cite{Bashinsky:2003tk}. This
phase shift in the acoustic oscillations of the CBR has been detected and it provides more evidence for only three neutrino
flavors~\cite{Follin:2015hya}.

\subsection{Galaxy and Structure Formation}

Cosmological neutrinos do not cluster like cold dark matter or baryons. Because of their extremely small masses, their velocity remains large enough to prevent them from falling into the gravitational wells created by dark matter on scales smaller than their effective {\index~Jeans scales}, thus leading to a suppression of the matter power-spectrum on scales smaller than the Jeans scales~\cite{Doroshkevich:1980vy,Bond:1980ha,Shafi:1984ek}. Mass fluctuations on scales smaller than these scales will thus grow at a smaller clustering rate than that expected~\cite{Lesgourgues:2006nd}. Their corresponding {\index~Jeans mass} is larger than the mass of the largest galaxies and can be on the scale of galaxy clusters. Observational indications of such large-scale clustering and implications for neutrino mass include gravitational lensing studies~\cite{Cooray:1999rv} and the number density of clusters of galaxies~\cite{Ichiki:2011ue,Colas:2019ret}.

Astronomical observations bearing on cosmological parameters $\Omega$ and $h$ play an important role in
neutrino physics and astrophysics. The power spectrum of the CMB is a function of the sum of the neutrino masses, $m_{\nu,tot}$ and other cosmological parameters. By comparing data from the Planck satellite with simulations of the development of structure in the universe the Planck Collaboration obtained an upper limit on the {\index~sum of the neutrino mass states} of $m_{\nu,tot} < 0.12$ eV~\cite{Planck:2018nkj,Planck:2018vyg}. 

A more detailed treatment of the dynamical effects of the cosmological neutrino background is beyond the scope of this short
treatment. For more extensive treatments reviewing the role of the C$\nu$B in cosmology, the reader is referred to
papers by Dolgov~\cite{Dolgov:2002wy} and Lesgourgues and Pastor~\cite{Lesgourgues:2006nd}.

\section{Direct Detection of the Cosmic Neutrino Background}

\subsection{Detection via the $Z$-boson resonance}

A potential scheme for detecting cosmological neutrinos was suggested via the use of the {\index~$Z$-boson resonance}~\cite{Weiler:1982qy}. The IceCube detector has observed the $W^-$ boson resonance~\cite{IceCube:2021rpz}, also known as the {\index~Glashow resonance}~\cite{Glashow:1960zz}, from neutrinos interacting with electrons in ice.
At the 6.3 PeV resonance energy $E_{\bar{\nu}_{e}} = M_W^2/2m_{e} = 6.3$ PeV,
electrons in the IceCube volume provide enhanced target cross sections for $\bar{\nu}_{e}$'s through the $W^-$ resonance channel via $\bar{\nu}_{e} + e^- \rightarrow W^- \rightarrow ~shower$. 

In addition to the Glashow resonance, the standard model predicts a $Z$-boson resonance via the neutral current interaction $\nu + \bar{\nu} \rightarrow Z \rightarrow ~shower$. The corresponding resonance energy is  $E_{\nu} = M_Z^2/2m_{\nu} \ge \sim 2 \times 10^{13}$ GeV, taking $m_{\nu} \le 0.2$ eV. In order to produce a neutrino of that energy, it would require the interaction of a proton with an energy an order of magnitude greater~\cite{Stecker:1978ah}. As there is no known mechanism for accelerating a proton to such an energy in an astronomical object, direct detection of cosmological neutrinos via the $Z$-boson resonance is doubtful.

\begin{figure}[h!]
\begin{center}
\includegraphics[width=0.9\textwidth]{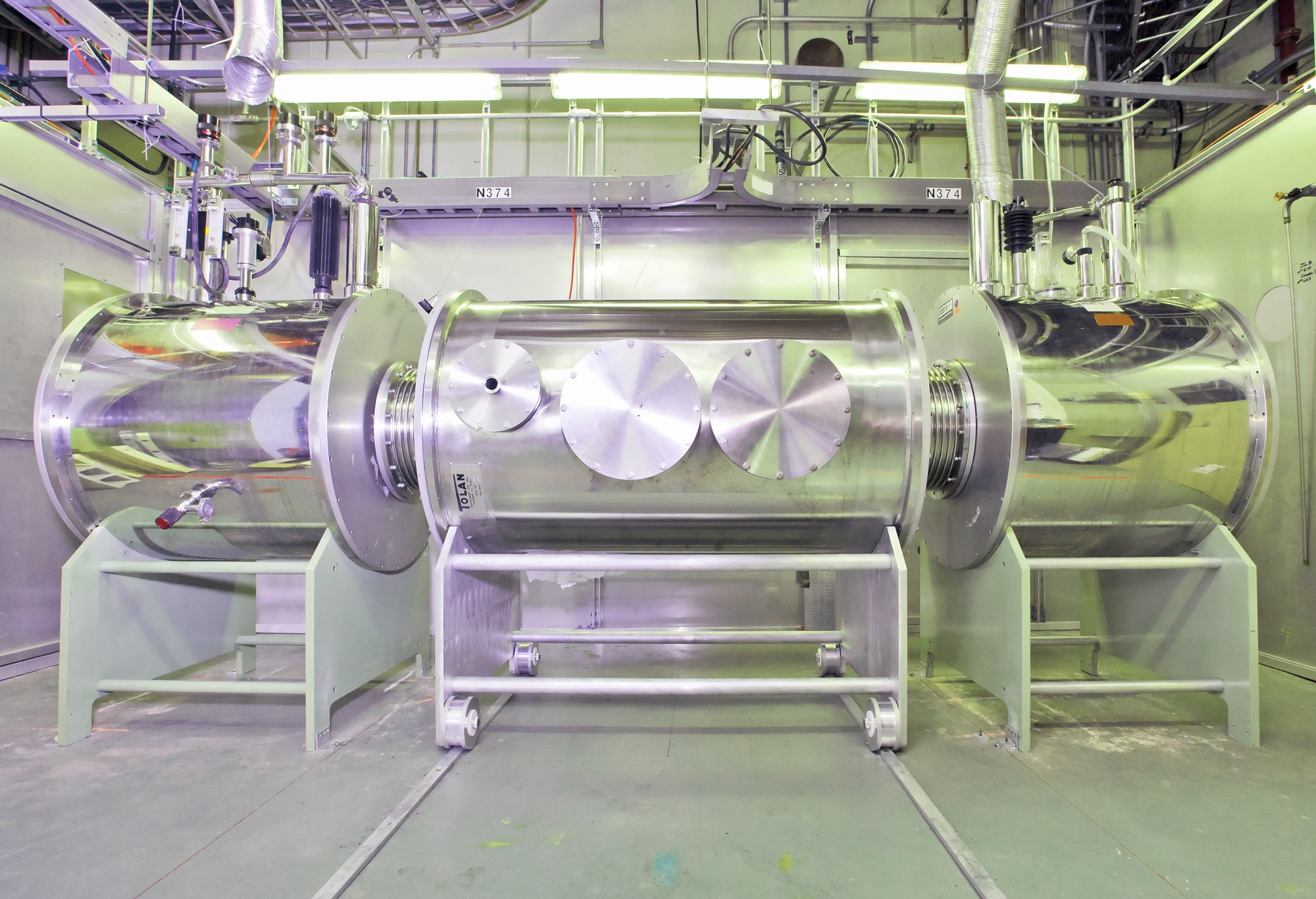}
\caption{The small-scale PTOLEMY prototype installed at the Princeton Plasma Physics Laboratory (February 2013).  Two horizontal bore NMR magnets are positioned on either side of a MAC-E filter vacuum tank.  The tritium target
plate is placed in the left magnet in a 3.35T field, and the RF tracking system is placed in a high uniformity 1.9T field
in the bore of the right magnet with a windowless APD detector and in-vacuum readout electronics. (Courtesy of the
{\index~PTOLEMY collaboration}.)
\label{fig:PTOLEMYphoto}}
\end{center}
\end{figure}

\subsection{Detection via neutrino capture}

As of this writing, there has been no direct detection of the very large number of very low energy cosmological
background. However, there is a proposed concept for such direct detection. The proposed experiment, called
PTOLOMY (Princeton Tritium Observatory for Light, Early-universe Massive-neutrino Yield)~\cite{Cocco:2007qv,Alvey:2021xmq}
~is based on detection via the neutrino capture processes on $\beta$-unstable {\index~tritium} nucleus
\begin{equation}
\nu_e + {^{3}\mathrm{H}} \rightarrow {^{3}\mathrm{He}} + e^{-}
 \label{eq:nu-capture-reaction}.
\end{equation}
The experimental signature of {\index~neutrino capture} is a peak in the electron spectrum that is displaced by $2 m_{\nu}$ above the {\index~{\it beta}-decay} endpoint. The signal would exceed the background from $\beta$-decay if the energy resolution is less than $0.7m_\nu $.  The capture rate depends on the nature of the neutrino mass. In principle, this provides a test that can distinguish between {\index~Dirac neutrinos} and {\index~Majorana neutrinos}. Assuming the neutrinos are non-relativistic, for a 100 g $^{3}\mathrm{He}$ target the capture rate for unclustered Dirac neutrinos is $\sim 4$ $yr^{-1}$ and twice that for Majorana neutrinos, taking {\index~helicity} into account~\cite{Long:2014zva}. At present there is not enough $^{3}\mathrm{H}$ available for the sensitivity needed, however, a prototype pathfinder experiment is in the works.

\section*{Acknowledgment} 
I would like to thank Sean Scully for helpful comments.
  
\newpage

\bibliographystyle{ws-rv-van}
\bibliography{Chapter3}
\end{document}